\documentclass[aps,pre,reprint,utf8, superscriptaddress, floatfix]{revtex4-1} 
\usepackage[T1]{fontenc} 
\usepackage[utf8]{inputenc}
\usepackage{graphicx} 
\graphicspath{{./images/}} 
\usepackage{xr} 
\usepackage{amsmath,amssymb}
\newcommand{\tonda}[1]{\left( #1 \right)}
\newcommand{\quadra}[1]{\left[ #1 \right]}

\usepackage{color}
\definecolor{crimson}{RGB}{220,20,60}
\definecolor{darkorchid}{rgb}{0.6, 0.2, 0.8}\definecolor{lightskyblue}{RGB}{135,206,250}
\definecolor{darkcyan}{rgb}{0, 0.5, 0.5}
\definecolor{lightcyan}{rgb}{0, 0.9, 0.9}
\definecolor{gold4}{RGB}{139,117,0}
\definecolor{lightgold1}{RGB}{238,220,130}

\begin{document}
\title{Power-law and log-normal avalanche size statistics in random  growth processes}
\author{Stefano Polizzi}
\affiliation{Ecole Normale Supérieure de Lyon, 69342, Lyon, France}
\affiliation{Univ. Bordeaux, CNRS, LOMA, UMR 5798, F-33405 Talence, France}
\author{Francisco-Jos{\'e} P{\'e}rez-Reche}
\affiliation{Institute for Complex Systems and Mathematical Biology, SUPA, University of Aberdeen AB24 3UE, United Kingdom}
\author{Alain Arneodo}
\affiliation{Univ. Bordeaux, CNRS, LOMA, UMR 5798, F-33405 Talence, France}
\author{Françoise Argoul}
\email{francoise.argoul@u-bordeaux.fr}
\affiliation{Univ. Bordeaux, CNRS, LOMA, UMR 5798, F-33405 Talence, France}

\begin{abstract}
We study the avalanche statistics observed in a minimal random growth model. The growth is governed by a reproduction rate  obeying a probability distribution with finite mean $\bar{a}$ and variance $v_a$. These two control parameters  determine if the avalanche size tends to a stationary distribution, (\emph{Finite Scale} statistics with finite mean and variance or \emph{Power-Law} tailed statistics with exponent $\in (1,3]$), or instead to a non-stationary regime with \emph{Log-Normal} statistics. 
 Numerical results and their statistical analysis are presented for a uniformly distributed growth rate, which are corroborated and generalized by mathematical results. The latter show that the numerically observed avalanche regimes exist for a wide family of growth rate distributions and provide a precise definition of the boundaries between the three regimes. 
\end{abstract}

\maketitle

In  complex systems with long-range spatio-temporal correlations avalanche processes are commonly observed.   Well-known examples of avalanches include the spreading of epidemics (or information) \citep{rhodes1997,pinto2011},  the price evolution of stock options in finance \citep{Black1973}, avalanches of neuron firings in the brain \citep{Beggs2003,friedman2012,roberts2014},  ``crackling noise'' exhibited by earthquakes \citep{gutenberg1942,gutenberg1956}, structural phase transitions~\citep{perez-reche_kinetics_2004,gallardo_avalanche_2010} and magnetic systems \citep{Sethna1993,sethna2001}, or avalanches of fractures in porous media~\citep{baro_statistical_2013} or living systems \citep{Polizzi2018,streppa2017a}. 
A crucial quantity to characterize avalanches is their size distribution, which allows theoretical and experimental results to be compared and can suggest mechanisms for the underlying avalanche dynamics. Notably, heavy-tailed distributions are often observed for avalanche size statistics and understanding them is important to determine the origin of the specific process.
From a practical viewpoint, it is often difficult to distinguish the type of heavy-tailed distributions on finite intervals, especially for limited size samples or noisy data. Pareto (power-law) and log-normal distributions are two of the most widely observed heavy-tailed distributions~\citep{kleiber2003,Sornette2006}. Many investigations have described heavy-tailed data in terms of Pareto or power-laws with exponential decays~\citep{Newman2005}. Careful statistical analyses, however, indicated that statistical evidence in support of a power-law distribution is often limited \citep{stumpf_critical_2012} and a log-normal distributions can often be a good alternative to describe heavy-tailed statistics~\citep{Clauset2009}. These difficulties are clearly exemplified by the ongoing controversy between log-normal and power-law distributions in neuroscience \citep{Beggs2012,Buzsaki2014} and complex networks \citep{Broido2019}. 
 The discrimination between power-law and log-normal distributions is even more challenging for data that can be modelled as a log-normal distribution at moderate sizes with a power-law tail  \citep{montroll1982}.
  
Several paradigmatic models have been proposed to explain the ubiquity of power-law avalanche size distributions. These include critical points in disordered systems \citep{Sethna1993,mehta2002,perez2003,perez2016,borja_da_rocha_rigidity-controlled_2020}, self-organized criticality (SOC) \citep{Bak1996}, marginal stability \citep{pazmandi_self-organized_1999} and, in more abstract terms, growth models \citep{yule1925} or branching processes \citep{Harris1963,Corral2013,gleeson2017,disanto_simple_2017}. In fact, some of these paradigms are related, or can even be mapped, to each other (e.g. SOC and branching processes \citep{zapperi1995} or branching processes and spin models \citep{handford_mechanisms_2013}).

Log-normal distributions are often explained in terms of stochastic multiplicative models of growth phenomena based on the law of proportionate effect. Here, we focus on Gibrat's process \citep{gibrat_1930,gibrat1931}, which can be viewed as a discrete time version of the so-called multiplicative noise \cite{sancho_analytical_1982}. Gibrat's process assumes that the size $z_i$ of an observable in generation $i$ grows proportionally to its size with a random reproduction (or growth) rate, $a_i$: $z_{i+1}=a_i z_i$. Assuming that the growth rates $\{a_i\}_{i=1}^{\infty}$ are independent random variables and the first two moments of $\ln a_i$ are finite for every $i$, the central limit theorem implies that $z_i$ is log-normally distributed for large $i$ \citep{Sornette2006}, or see  \cite{redner1990} for a more precise approximation. Avalanches are typically regarded as bursts of activity which in our case would correspond to excursions of $z$ that asymptotically return to the absorbing state with $z=0$ after being perturbed from this state. Since Gibrat's variable $z$ can either approach zero or $\infty$ when iterated, we extend the usual avalanche definition to encompass the case in which $z$ does not return to $0$ but grows indefinitely, as in supercritical branching processes~\citep{Harris1963}. The size of an avalanche corresponds to the sum of $z_i$ over generations.  Despite the fact that the distribution of $z_i$ is reasonably well understood for Gibrat's processes, little is known about the avalanche size distribution. In general, a log-normal distribution for $z_i$ does not imply a log-normal distribution for  the avalanche size and Gibrat's process cannot be regarded as an explanation of log-normal avalanche size statistics.

Here, we push further the comparative analysis between power-law and log-normal distributions by studying the avalanche size distribution of Gibrat's processes. By means of mathematical results and numerical simulation examples, we reveal rich avalanche behavior which, in particular, includes power-law and log-normal avalanche size statistics.

 
\paragraph{The model.--} The basis of our avalanche model is the following multiplicative process: 
\begin{equation}
z_{i+1} = a_i z_i = a_i z_0 \prod ^{i-1}_{j=0}a_j~. 
\label{eq:process_def}
\end{equation}
The initial value $z_0$ for the process represents a perturbation of the system from the $z=0$ absorbing state. We set $z_0=1$ (a different  positive value of $z_0$ would only lead to a time shift). The 
reproduction rates $\{a_n\}_{n=0}^{\infty}$ are independent and identically distributed (i.i.d.) random variables with finite
 mean $\bar{a}=\mathbb{E}_a[a]$ and variance $v_a=\mathbb{E}_a[a^2]-\mathbb{E}_a^2[a]$. In our model, the probability density function (PDF) for the reproduction rate, $f_a(a)$, is required to have a non-negative support to ensure that $z_i \ge 0$ at every generation $i$. 
 
The avalanche size after $T$ generations is given by the following sum:
\begin{equation}
Z_T = \sum_{i = 1}^{T} z_i =a_1 + a_1 a_2 + \dots + a_1 \dots  a_T~.
\label{eq:Z_sum}
\end{equation}

 Our aim is to understand the dependence of the PDF for the avalanche size, $p(Z_T)$,  on  the two parameters of the reproduction rate distribution, $\bar{a}$ and $v_a$. Eq.~\eqref{eq:Z_sum} shows that the avalanche size $Z_T$ is a random variable given by the sum of $T$ random variables. The challenge in calculating $p(Z_T)$ is that $\{z_i\}_{i=1}^T$ are correlated and the Central Limit Theorem does not apply in general \citep{grimmett_probability_2001}. Accordingly, there is no reason to expect that $Z_T$ is normally distributed for large $T$ as one would expect if $\{z_i\}_{i=1}^T$ were uncorrelated. In fact, Eq.~\eqref{eq:Z_sum} shows that $Z_T$ is a Kesten scalar variable~\citep{kesten_random_1973}. Within this context, power-law tails have been reported for $p(Z_T)$ under quite general conditions~\citep{kesten_random_1973,goldie_implicit_1991,calan_distribution_1985,gautie_matrix_2021}. Here, we identify power-law decay as one of three generic behaviours for $p(Z_T)$. In addition, our analysis establishes a conceptual link between a Kesten recursion and the size of avalanches described as a Gibrat's multiplicative process.

We first present results of numerical simulations for a specific PDF $f_a(a)$ that show the existence of three different regimes for $p(Z_T)$. After that, we mathematically demonstrate that the numerically observed avalanche regimes are expected for any $f_a(a)$ with finite first and second moments for $a$ and $\ln a$.

\paragraph{Numerical results.--}  Here, we present results for a uniformly distributed reproduction rate, $a_i \sim \mathcal{U}(b,c)$, with $0 \leq b < c$. The uniform distribution is a simple and flexible choice that allows the dependence of $p(Z_T)$ on $\bar{a}$ and $v_a$ to be systematically studied by independently tuning the parameters $\bar{a} = (b+c)/2$ and $v_a= (c-b)^2/12$.  In \citep{Supp2021}, we present qualitatively similar results for exponentially (Fig.~S2 of Sec.~IV)  and Poisson (Fig.~S3 of Sec.~V) distributed growth rates. In both cases, however, $\bar{a}$ and $v_a$ cannot be independently tuned.

\begin{figure}
\centering
\includegraphics[width = \columnwidth]{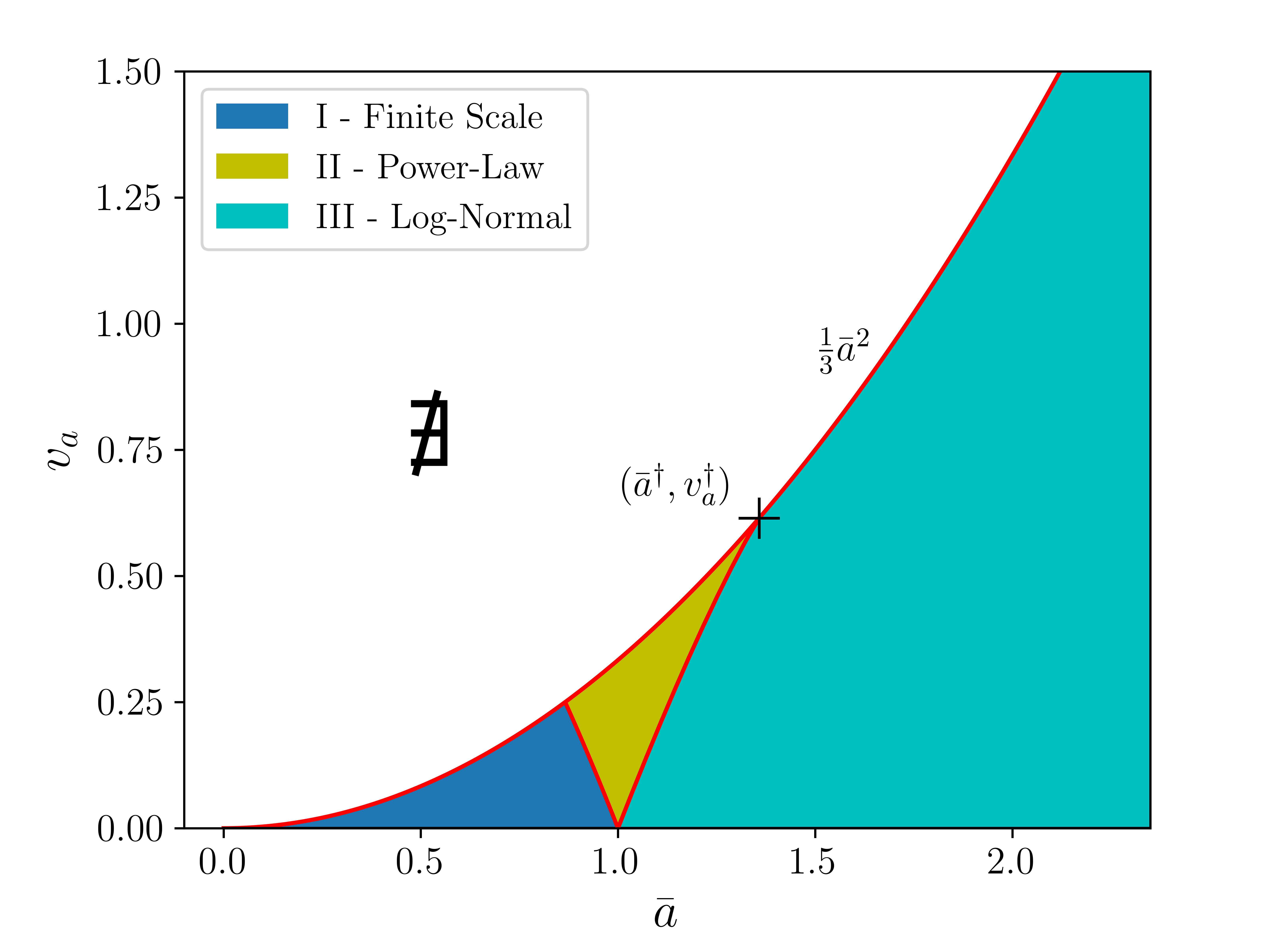}
\caption{Phase diagram on the $(\bar{a},v_a)$ space showing three regimes for the avalanche size distribution for uniformly distributed growth rate, $a \sim \mathcal{U}(b,c)$. In regime I (Finite Scale), $p(Z_T)$ converges to an asymptotic PDF $p(Z)$ with finite mean and variance. In regime II,  $p(Z_T)$ converges to  an asymptotic $p(Z)$ with a power-law tail. In regime III, $Z_T$ is non-stationary and $p(Z_T)$ approaches a log-normal distribution for large $T$, with $T$-dependent parameters. All regimes are bounded from above by the condition $v_a \leq \bar{a}^2/3$ ensuring $b>0$ and $(\bar{a}^{\dag} \simeq 1.36, v_a^{\dag} \simeq 0.61)$. Boundaries between different regimes were analytically obtained. 
\label{fig:schema}}
\end{figure}

Fig.~\ref{fig:schema} shows  three avalanche regimes identified for  uniformly distributed $a$ on the $(\bar{a}, v_a)$ space. The region of the space where a random growth processes is possible depends on the specific PDF for the growth rate. For uniformly distributed $a$, the region is restricted to $\bar{a} \geq 0$ and $v_a \in [0,\bar{a}^2/3]$. For a given $\bar{a}$, the upper bound for $v_a$ reflects the constraint $b \geq 0$. For a general $f_a$, the upper bound is given by the condition $a \geq 0$.
 
\begin{figure}[t]
\includegraphics[width = \columnwidth]{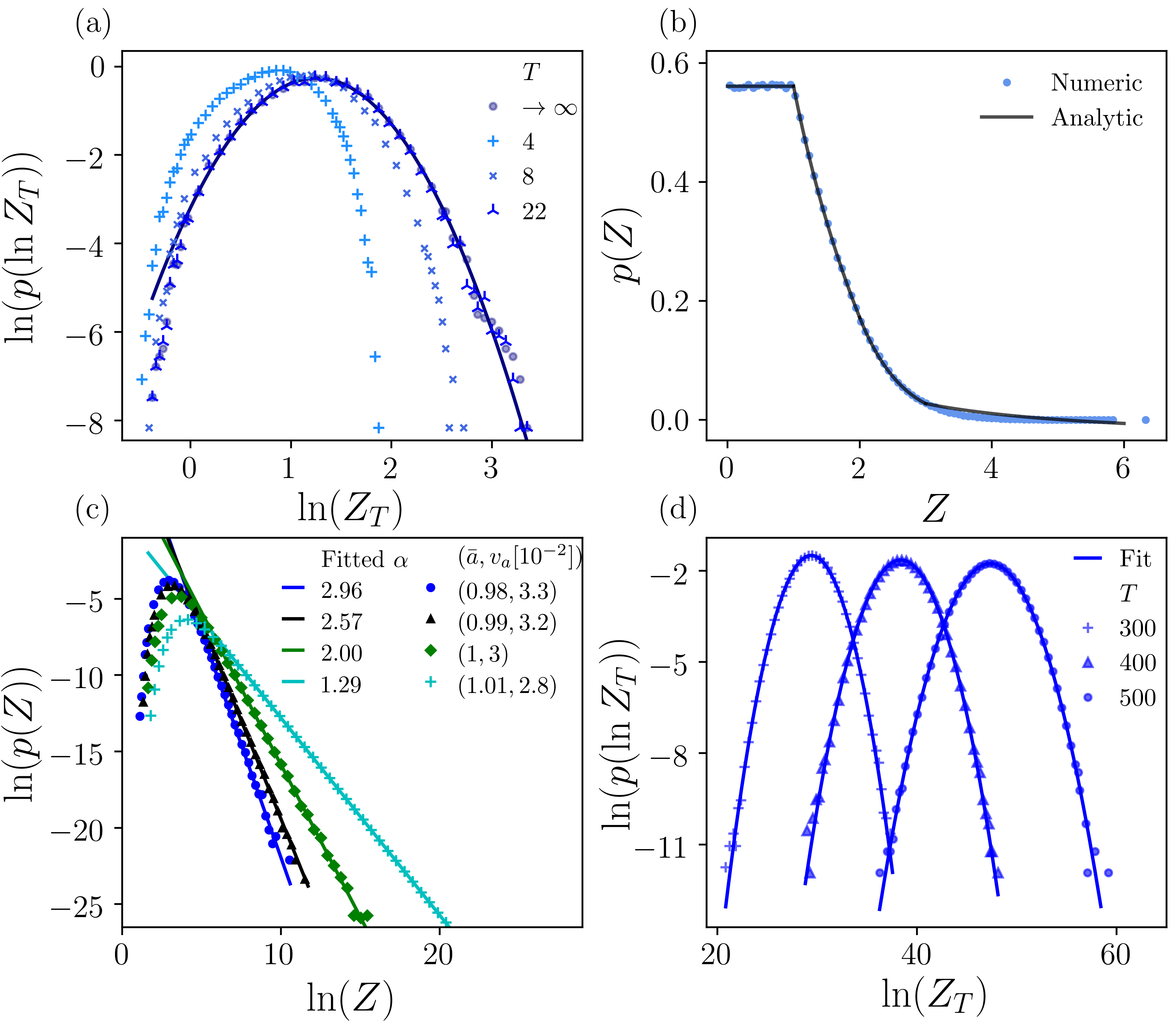}%
\caption{Examples of avalanche size PDFs for a uniformly distributed growth rate, $a \sim \mathcal{U}(b,c)$. (a) Regime I: convergence of the PDF of $\ln Z_T$ towards an asymptotic distribution with $T$. $(\bar{a},v_a)=(0.8,0.06)$, solid line: fit of a log-normal distribution to the asymptotic distribution. (b) Asymptotic PDF $p(Z)$ for $(\bar{a}, v_a) = (0.5,\bar{a}^2/3)$ (i.e. $b=0$ corresponding to the upper bound for $v_a$ in Fig.~\ref{fig:schema}). Solid line: analytical solution. (c) Regime II: asymptotic $p(Z)$ in log-log scale. Solid lines:  fits of power-laws with exponents $\alpha$ given in the legend. (d) Regime III: logarithm of $p(\ln(Z_T))$  with $(\bar{a},v_a) = (1.1,0.013)$.  Solid lines:  fits of a log-normal distribution to the data for each given $T$. Symbols (respectively solid lines) are used for numerical results (resp. maximum likelihood fits or analytical results).} 
 \label{fig:numerical_results}
\end{figure}

Regime I (dark blue region in Fig.~\ref{fig:schema}) is characterized by avalanches for which $z_i$ approaches zero after a finite number of generations in such a way that the mean and variance of $p(Z_T)$ are finite for every $T$. This regime is referred to as the \emph{Finite Scale} regime, as opposed to \emph{scale-free} distributions which lack of a typical scale. Below, we mathematically show that the necessary condition for the first two moments of $p(Z_T)$ to be finite is $v_a+\bar{a}^2<1$, for any $f_a(a)$. In particular, this condition defines the boundary between regions I and II shown in Fig.~\ref{fig:schema} for a uniformly distributed $a$. In regime I, $p(Z_T)$ converges to an asymptotic PDF, $p(Z)$, after a finite number of generations, $T_z$. Fig.~\ref{fig:numerical_results}(a) shows an example of the convergence of $p(Z_T)$ to  $p(Z)$ after $T_z = 22$ generations. The rate of convergence decreases as the boundary with region II is approached. The specific shape of $p(Z)$ depends on $\bar{a}$ and $v_a$. Phenomenologically, we observe that  
 the subset of  region I with $v_a \lesssim \bar{a}-0.7 $ (then excluding values of $v_a$ just below the upper boundary) shows  distributions that are compatible with a log-normal (see Fig.~\ref{fig:numerical_results}(a)). This result is reminiscent of cases in which 
a log-normal distribution was observed as the asymptotic  distribution for the sum of a large but finite number of uncorrelated and log-normal, or, more generally, positively skewed  random variables \citep{da_costa2000, Mouri2013}. The comparison of our results with those in \citep{da_costa2000, Mouri2013}, however, is not complete due to  the presence of correlations between the random variables $\{z_i\}_{i=1}^T$, defining $Z_T$ in our model. In fact, a log-normal like distribution is observed in a good part of region I in the phase diagram, but it is not the only possible shape for $p(Z)$ in this regime. For instance,  Fig.~\ref{fig:numerical_results}(b) shows an example of $p(Z)$ observed at a point along the upper bound for region I in the phase diagram (line with $v_a=\bar{a}^2/3$ in Fig.~\ref{fig:schema}).  See Sec.~III~A of \citep{Supp2021} for more examples of $p(Z)$ in this regime.

In regime II, $p(Z_T)$ converges to an asymptotic PDF $p(Z)$ with a power-law (or Pareto) tail   $Z^{-\alpha}$ (see Fig.~\ref{fig:numerical_results}(c)). A maximum likelihood fit to the data (Sec.~VI of \citep{Supp2021}) reveals that the exponent $\alpha$ takes values that range from $\alpha = 3$ at the boundary with regime I to $\alpha = 1$ at boundary with regime III. Below we mathematically show that this range for $\alpha$ holds beyond the uniformly distributed $a$ used for the simulations shown in Fig.~\ref{fig:numerical_results}.

In regime III, avalanches grow indefinitely and $p(Z_T)$ does not converge to a $T$-independent PDF. Instead, the location and spread of $p(Z_T)$ monotonically increase with $T$ (Fig.~\ref{fig:numerical_results}(d)). Interestingly, $p(Z_T)$ can be very well described by a log-normal distribution with $T$-dependent parameters. This is corroborated by a likelihood ratio test \citep{Alstott2014} and parametric bootstrap \citep{davison1997} (see more details in Sec.~VI of \citep{Supp2021}).

\paragraph{Mathematical results.--} We now show mathematically that the three avalanche regimes illustrated numerically for uniformly distributed $\{a_i\}$ can be observed for generic distributions $f_a(a)$ with non-negative support and finite first and second moments for $a$ and $\ln a$. This analysis also provides general conditions satisfied at the boundaries between different regimes.

To study the PDF of the avalanche size $Z_T$ for a generic $f_a(a)$, we express  Eq.~\eqref{eq:Z_sum} as $Z_T=a_1\tonda{1+X_T}$. Here, $X_T=\sum_{i=2}^T \prod_{j=2}^i a_j$ is a random variable whose behavior at large $T$ determines whether the system is in regime I, II or III. Regime III corresponds to situations in which $z_{T+1}/a_1 = \prod_{j=2}^T a_j$ increases monotonically with $T$. In this case, an infinite avalanche occurs in which $z$ grows indefinitely and $Z_T$ obeys a non-stationary log-normal distribution for large $T$, provided $\mathbb{E}_a[\ln^2 a]<\infty$. 
Indeed, in this case $X_T$ is distributed as $Z_{T-1}$ and therefore $Z_T \sim a_1 Z_{T-1}$ for large $T$. We then conclude that $Z_T$ is given by the product of $T$ i.i.d.  positive random variables obeying $f_a(a)$  and, provided $\mathbb{E}_a[\ln^2 a]<\infty$,  $Z_T$ obeys a log-normal distribution with expectation and variance that increase exponentially with $T$ (see expressions for $\mathbb{E}[Z_T]$ and $\mathrm{Var}[Z_T]$ in Sec.~II of \citep{Supp2021}). In other words, $Z_T$ essentially obeys Gibrat's law in regime III. 

Regimes I and II are observed when the product $\prod_{j=2}^T a_j$ tends to zero for large $T$ and therefore $z$ asymptotically approaches the absorbing state with $z=0$. In this sense, regimes I and II define the absorbing phase of the model.  Under this condition, $X_T \sim Z_T$ and therefore $Z_T$ tends to a stationary random variable $Z$ with $p(Z)$ given by the following equation (see a derivation in Sec.~I of \citep{Supp2021}):
\begin{equation}
p(Z)=  \mathbb{E}_a \left[a^{-1} p \left(\frac{Z}{a}-1 \right) \right]~,
\label{eq:pZ}
\end{equation}

This can be reduced to a homogeneous Fredholm integral equation of the second kind \citep{press_numerical_2007} that is difficult to solve in general. We only solved it analytically for a specific case with $a \sim \mathcal{U}(0,c)$ which accurately matches the numerical results in region I, as shown in Fig.~\ref{fig:numerical_results}(b) \citep{Supp2021} (see \citep{calan_distribution_1985} for other exact solutions of Eq.~\eqref{eq:pZ}). Even if Eq.~\eqref{eq:pZ} cannot be analytically solved in general,  it is easy to show that regime I, where the first two moments of $Z$ are finite, is observed for any distribution $f_a(a)$ provided $\mathbb{E}_a[a]=\bar{a}<1$ and $\mathbb{E}_a[a^2]=v_a+\bar{a}^2<1$ \citep{Supp2021} (Sec.~II). The boundary between regimes I and II is then  given by the condition $\mathbb{E}_a\quadra{a^2}=1$, or equivalently $v_a=1-\bar{a}^2$ for any PDF $f_a(a)$. 

To investigate the properties of regime II and its boundary with regime III, we insert a power-law tail ansatz, $p(Z) \propto Z^{-\alpha}$, into Eq.~\eqref{eq:pZ}. From this we find that the exponent $\alpha$ is given by the zeros of the function
\begin{equation}
h(\alpha)=  \mathbb{E}_a[a^{\alpha-1}]-1~.  
\label{eq:halpha}
\end{equation}

The function $h(\alpha)$ has a root at $\alpha=1$ due to the normalization of $f_a(a)$ which implies $\mathbb{E}_a[1]=1$. However, we are only interested in roots with $\alpha \in (1,3]$, irrespective of the specific form of $f_a(a)$. The condition $\alpha >1$ ensures that  $p(Z)$ is normalizable and the condition $\alpha \leq 3$ corresponds to the boundary between regimes I and II where $\mathbb{E}_a \quadra{a^2}=1$. 

\begin{figure}[t]
\includegraphics[width = \columnwidth]{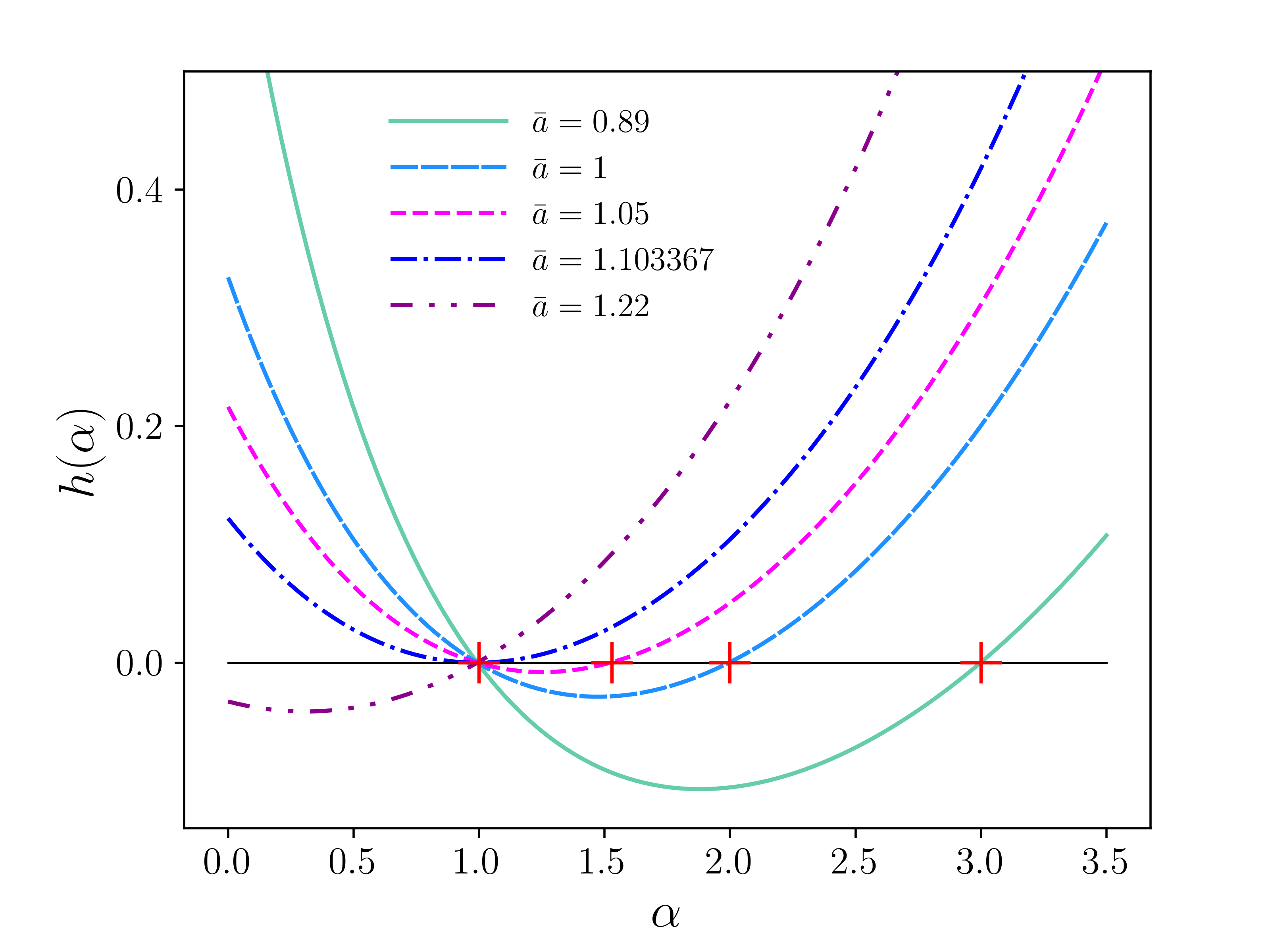}
\caption{Function $h(\alpha)$ (Eq.~\eqref{eq:halpha}) for a uniformly distributed reproduction rate (see Sec.~III~C of  \citep{Supp2021} for the exact analytic form). Different curves correspond to different values of $\bar{a}$ for growth processes with $v_a=0.2$. The exponent of the power-law tail for $p(Z)$ in regime II corresponds to the roots of $h(\alpha)$ in the interval of $\alpha \in (1,3]$. This corresponds to curves with $0.89 \lesssim \bar{a} \lesssim 1.1$ in this example.  
 \label{fig:critical_line}}
\end{figure}

Fig.~\ref{fig:critical_line} illustrates the behavior of $h(\alpha)$ for a uniformly distributed $a$ with fixed $v_a$ and various values of $\bar{a}$. A similar behavior is expected for any distribution $f_a(a)$ since $h(\alpha)$ is strictly convex for any $f_a(a)$ in the interval with $\alpha \geq 1$. Therefore, $h(\alpha)$ has at most one minimum and one root in the interval of interest, $(1,3]$. As illustrated in Fig.~\ref{fig:critical_line}, the root of $h(\alpha)$ decreases with increasing $\bar{a}$ from the value $\alpha=3$ at the boundary between regimes I and II to approach the minimum admissible value, $\alpha = 1$, which marks the transition from regime II to regime III. At the transition between regime II and III, the minimum of $h(\alpha)$ occurs at $\alpha=1$ and this leads to the condition 
\begin{equation}
    h'(1)=\mathbb{E}_a[\ln a]=0
    \label{eq:Cond_boundary_II_III}
\end{equation} 
for the boundary between the two regimes. The specific shape of the boundary in the space $(\bar{a},v_a)$ depends on the specific distribution of $a$. Eq.~\eqref{eq:Cond_boundary_II_III} allows the relation between $\bar{a}$ and $v_a$ to be obtained for any $f_a(a)$. In particular, we obtained analytical results for uniformly and exponentially distributed $a$ which compare well with numerical results (see Fig.~\ref{fig:schema} and more details in \citep{Supp2021}).
 
In fact, the condition $\mathbb{E}_a[\ln a]=0$ holds at the boundary between regimes with stationary and non-stationary $p(Z_T)$, irrespective of the power-law assumption made for regime II. Indeed, if $\mathbb{E}_a[\ln^2 a]<\infty$, the strong law of large numbers  \citep{grimmett_probability_2001} allows us to express  $\prod_{j=2}^T a_j$ as $e^{T \mathbb{E}_a[\ln a]}$ for large $T$. Accordingly, the sign of $\mathbb{E}_a[\ln a]$ determines whether $\prod_{j=2}^T a_j$ tends to zero and $Z_T$ reaches a stationary regime (if $\mathbb{E}_a[\ln a]<0$, regimes I and II) or increases for increasing $T$ and $Z_T$ is not stationary (if $\mathbb{E}_a[\ln a]>0$, regime III). For a power-law $p(Z)$, one can see the change in sign of $\mathbb{E}_a[\ln a]$ at the transition between regimes II and III in terms of the slope $h'(1)$ which is negative in regime II and positive in regime III (see Fig.~\ref{fig:critical_line}).

\paragraph{Conclusions.--}  We showed that  power-law \citep{takayasu1997,Newman2005} and log-normal avalanches  can coexist, in a minimal random growth model with a reproduction rate with finite mean and variance. Interestingly, the power-law tail exponent $\alpha$ can be  continuously tuned in the range $(1,3]$, by varying the control parameters. 
Therefore, our study can explain several power-laws found in natural or human processes, such as the ones described in \citep{Newman2005}, whose exponents are also almost always in the interval $(1,3]$. Many of these phenomena have an underlying multiplicative process and can be interpreted as avalanches.  
We have focused on growth processes with finite value for $\bar{a}$ and $v_a$. It is worth noting, however, that the condition determining a transition from a stationary distribution to a non-stationary one (Eq.~\eqref{eq:Cond_boundary_II_III}) and the definition of the exponent $\alpha$ (Eq.~\eqref{eq:halpha}) remain valid even if one (or both) of the parameters diverges (provided $\mathbb{E}_a[\ln^2 a]< \infty$). This is consistent with previous studies, where a power-law distributed growth rate was considered \citep{saichev2005}.  

The model studied in this paper can be seen as a generalization of branching processes which correspond to a specific distribution for the growth rate (Sec.~VII of \citep{Supp2021}). In particular, the exponent $\alpha=3/2$ observed for power-law distributed avalanche sizes in critical branching processes \citep{Harris1963,Corral2013} is contained within the interval $(1,3]$ obtained here. Besides power-law and
log-normal distributions, we observed, especially in the Finite Scale regime, but also along the upper bound of Fig.~\ref{fig:schema}, less common distributions for avalanches, but nonetheless observed experimentally, such as the bimodal shape shown in Sec. III~A of Supp Mat.

We assumed that $\bar{a}$ and $v_a$ remain constant during the course of the avalanches. One could, however, consider dynamical parameters to mimic  feedback mechanisms such as vaccination in epidemics or refractoriness in neuronal avalanches. In this case, our phase diagram in Fig.~\ref{fig:schema} can be used to propose qualitative scenarii for the ongoing controversy on log-normal or power-law distributions in neuroscience and other domains \citep{Beggs2012,Buzsaki2014,Broido2019}. Indeed, besides giving an interpretation of the different distributions in terms of $\bar{a}$ and $v_a$, it has to be seen as a guide for avalanche distributions, even for more realistic situations where the control parameters are functions of time, as in \citep{Polizzi2018, Polizzi2021,handford_mechanisms_2013,perez2016}. This corresponds to a path in the diagram, where the distributions are combined with different weights. For example an avalanche with initial parameter values in region III, shifting in time toward region II or I (because of external feedbacks as refractoriness in the brain), would give a log-normal dominating distribution. This qualitative scheme suggests that the three avalanche regimes identified here are relevant to realistically complex situations with non-stationary $a_t$. A more precise description of the avalanche size in such situations, however, would require extending our analysis to Gibrat's processes with non-stationary $a_t$.

\begin{acknowledgments}
We thank J.P. Bouchaud for constructive comments. We acknowledge financial support from the Agence Nationale de la Recherche (ANR grant number ANR-18-CE45-0012-01)  and from the French Research Ministry (MESR) (contract No. 2017-SG-D-09) and from ENS Lyon for SP PhD funding. FJPR acknowledges financial support from the Carnegie Trust. 
\end{acknowledgments}

\bibliographystyle{apsrev4-1} 

\end{document}


\title{Power-law and log-normal avalanche size statistics in random  growth processes - Supplementary material}
\author{Stefano Polizzi}
\email{stefano.polizzi@u-bordeaux.fr}
\affiliation{Ecole Normale Supérieure de Lyon, 69342, Lyon, France}
\affiliation{Univ. Bordeaux, CNRS, LOMA, UMR 5798, F-33405 Talence, France}
\author{Francisco-José Pérez-Reche}
\affiliation{Institute for Complex Systems and Mathematical Biology, SUPA, University of Aberdeen AB24 3UE, United Kingdom}
\author{Françoise Argoul}
\affiliation{Univ. Bordeaux, CNRS, LOMA, UMR 5798, F-33405 Talence, France}

\maketitle

\section{Derivation of Eq. (3)of the main text}
\label{sec:derivatio_eq3}

Eq.~\eqref{eq:pZ} of the main text holds for cases in which $\lim_{T \to \infty}\prod_{j=2}^T a_j \rightarrow 0$ so that $Z_T$ tends to a stationary random variable $Z=a(1+X)$, where $X=\lim_{T\to \infty} \sum_{i=2}^T \prod_{j=2}^i a_j$ is distributed as $Z$. Since $a$ and $X$ are independent random variables, $p(Z)$ can be obtained from the distributions of $X$ and $a$ as follows: 
\begin{equation}
\begin{gathered}
    p(Z)=\int_0^{\infty} \text{d}a f_a(a) \int_{0}^{\infty} \text{d}x p(x) \delta(Z-a(1+x))=\\
    =\int_0^{\infty} \text{d}a f_a(a) a^{-1} \int_{0}^{\infty} \text{d}x p(x) \delta\left(x-(Z/a-1)\right)=\\
    =\int_0^{\infty} \text{d}a f_a(a) a^{-1} p(Z/a-1)= \mathbb{E}_a \left[a^{-1} p \left(\frac{Z}{a}-1 \right) \right]~.
\end{gathered}
\end{equation}
Here, we used the property of the Dirac-$\delta$ distribution: $\delta (a x) = \frac{1}{|a|} \delta(x)$.
This equation is a Fredholm Integral Equation and, if solved, defines the full shape of the PDF $p(Z)$ for any choice of the reproduction rate distribution. For instance it will be used in the following Sections \ref{sec:analytic_unif}, \ref{sec:analytic_b=0}, \ref{sec:analytic_exp} for uniform or exponential distributed reproduction rates. 

\section{Expectation and variance of $Z_T$}
\label{sec:analytic_exp_and_var}

In this section we give expressions for the expected value $\mathbb{E}[Z_T]$ and variance $\mathrm{Var}[Z_T]$ of $Z_T$ which are valid for any $f_a(a)$.  

Let us denote the second moment of the reproduction rate as $m_2 = \mathbb{E}_a [a^2]$. If $\bar{a}$ and $m_2$ are both either larger or smaller than $1$, the calculation of $\mathbb{E}[Z_T]$ and $\mathrm{Var}[Z_T]$ reduces to summations of geometric series and, from Eq.~\eqref{eq:Z_sum} of the main text, one obtains:   
\begin{equation}
\mathbb{E}[Z_T] = \sum_{i=1}^T \tonda{\mathbb{E}\quadra{a}}^i = \frac{\bar{a} - \bar{a}^{T+1}}{1-\bar{a}},
\label{eq:averageZ_case5}
\end{equation}
and, with some more steps:   
\begin{equation}
\label{eq:varZ_case5}
\begin{gathered}
\mathrm{Var}[Z_T] = \sum_{j=1}^T m_2^j + 2\sum_{j=1}^{T-1}\sum_{k=1}^{T-j} m_2^j\tonda{\mathbb{E}[a]}^k - \mathbb{E}[Z]^2 = \frac{m_2- m_2^{T+1}}{1-m_2} + 2 \sum_{j=1}^{T-1} m_2^j\tonda{\frac{\bar{a} - \bar{a}^{T-j+1}}{1-\bar{a}}}- \mathbb{E}[Z]^2= \\ =
 2 \frac{\bar{a}}{1-\bar{a}}\quadra{\frac{m_2 - m_2^T}{1-m_2} - \frac{\bar{a}^{T+1}}{\bar{a}-m_2}  \tonda{\frac{m_2}{\bar{a}} - \frac{m_2^T}{\bar{a}^T} }}  +\frac{m_2 - m_2^{T+1}}{1-m_2} - \tonda{\frac{\bar{a}-\bar{a}^{T+1}}{1-\bar{a}}}^2.
\end{gathered}
\end{equation}
  We can easily see that if $m_2 < 1$ (and therefore $v_a<1 - \bar{a}^2$ and $\bar{a} < 1$), both $\mathbb{E}[Z_T]$ and $\mathrm{Var}[Z_T]$ converge to the finite values: $ \mathbb{E}[Z] = \bar{a}/(1-\bar{a})$ and $\mathrm{Var}(Z) =m_2/(1-m_2)\tonda{1+2 \bar{a}/(1-\bar{a})} - \bar{a}^2/(1-\bar{a})^2$.   
This is why we defined the distributions belonging to region I as Finite Scale. With note that these expressions for $\mathbb{E}[Z]$ and $\mathrm{Var}[Z]$ can also be obtained by integration from the formal expression of $p(Z)$ given in Eq.~\eqref{eq:pZ} of the main text. 

When $m_2 = 1$ (implying $\bar{a}<1$), the computation of $\mathrm{Var}[Z_T]$ results in:  
\begin{equation}
\mathrm{Var}(Z_T) = T + 2 \tonda{T-1}\frac{\bar{a}}{1-\bar{a}} -\frac{2 \bar{a}^{2}}{1-\bar{a}} \tonda{\frac{\bar{a}^{T-1}-1}{\bar{a} -1}}- \tonda{\frac{\bar{a} - \bar{a}^{T+1}}{1-\bar{a}}}^2, \label{varZ_case2}
\end{equation}
which in the large $T$ limit shows a linear divergence with $T$: $\mathrm{Var}(Z_T) \simeq T + 2 \tonda{T-1}\bar{a}/(1-\bar{a})$. The expected value  is still convergent, since $\bar{a} <1$.

As long as $m_2 > 1$ and $\bar{a}<1$, $\mathrm{Var}(Z_T)\sim m_2^T$ for large $T$ and it diverges exponentially with $T$ (from Eq.~\eqref{eq:varZ_case5}). In contrast, $\mathbb{E}[Z_T]$ converges as in the Finite Scale region. 

For $\bar{a} = 1$, the expectation grows linearly with $T$:
\begin{equation} 
\mathbb{E}[Z_T] = T,\label{eq:averageZ_case4}
\end{equation}
 while the variance has a more complicated expression, still showing an asymptotic exponential divergence with $T$, given by:
\begin{equation}
\begin{gathered}
\mathrm{Var}(Z_T) = \frac{1}{\tonda{1-m_2}^2} (m_2^{T+2} + m_2^{T+1}  - m_2^2\tonda{1+T}^2  + m_2\tonda{2 T^2 + 2T -1} - T^2).\label{eq:varZ_case4}
\end{gathered}
\end{equation}    
This result is based on the calculation of the following series: $$s_T = \sum_{j=1}^T j m_2^j = \frac{m_2 - m_2^{T+1}(1+T(1-m_2))}{(1-m_2)}^2.$$

Finally, cases with  $\bar{a}>1$ and $m_2 >1$ partially correspond to regime III, where Eqs.~\eqref{eq:averageZ_case5} and \eqref{eq:varZ_case5} are valid and predict an exponential divergence of  $\mathbb{E}[Z_T]$ and $\mathrm{Var}(Z_T)$ with $T$.
\begin{equation}
\begin{gathered}
\mathbb{E}[Z_T] \simeq \frac{\bar{a}^{T+1}}{\tonda{\bar{a} - 1}},\\
\mathrm{Var}(Z_T) \simeq \frac{2 \bar{a}\bar{a}^{T+1} m_2^T}{\quadra{\tonda{1-\bar{a}}\tonda{\bar{a} -m_2} \bar{a}^T}}. 
\end{gathered}
\label{eq:m_Z_s_Z_case5}
\end{equation}
Since we have numerical and statistical evidence that in region III the PDF of $Z_T$ is a log-normal, we can define the parameters $\mu_Z$ and $\sigma_Z$ such that $p(Z) \sim \text{Log-Normal}(\mu_Z, \sigma_Z)$. Then in the large $T$ limit we have:

\begin{equation}
\begin{gathered}
\mu_Z \simeq \ln\tonda{\bar{a}^{T+1}/\tonda{\bar{a} - 1}} - \frac{1}{2}\ln\tonda{\frac{2 \tilde{a}^T (\bar{a} -1 )}{\tonda{\tilde{a} - \bar{a}}\bar{a}^{2T}} + 1}, \\
\sigma_Z^2 \simeq \ln\tonda{\frac{2 \tilde{a}^T (\bar{a} -1 )}{\tonda{\tilde{a} - \bar{a}}\bar{a}^{2T}} + 1}. \label{sigmaZ_if_lognormal_case_5}
\end{gathered}
\end{equation}
From this we can define a standardized variable $t=(\ln Z_T-\mu_Z)/\sigma_Z$ which is stable and does not depend on $T$, since both $\mu_Z$ and $\sigma_Z^2$ depend linearly on $T$, as for the Central Limit Theorem. 
Note, however, that the region in the parameter space leading to exponential divergence of both $\mathbb{E}[Z_T]$ and $\mathrm{Var}(Z_T)$ with time (\textit{i.e.} $\bar{a}>1$ and $m_2 >1$) is not equivalent to region III, since it is the case also for the part of region II where $\bar{a} > 1$ (see for instance cyan crosses in Fig.~\ref{fig:numerical_results}(c)). 

To summarize there are $5$ different asymptotic behaviors of the expectation and the variance of $Z_T$:

\begin{itemize}
\item[1 $-$] If $\bar{a} <1$ and $m_2 < 1$, $\mathbb{E}[Z_T]$ and $\mathrm{Var}(Z_T)$  converge;
\item[2 $-$] if $m_2 = 1$, $\mathbb{E}[Z_T]$ converges and $\mathrm{Var}(Z_T)$ diverges linearly;
\item[3 $-$] if $m_2 > 1$ and $\bar{a} < 1$, $\mathbb{E}[Z_T]$ converges and $\mathrm{Var}(Z_T)$ diverges exponentially;
\item[4 $-$] if  $\bar{a} = 1$, $\mathbb{E}[Z_T]$ diverges linearly with $T$ and $\mathrm{Var}(Z_T)$ diverges exponentially;
\item[5 $-$] if $\bar{a} > 1$, both $\mathbb{E}[Z_T]$ and $\mathrm{Var}(Z_T)$  diverge exponentially.
\end{itemize} 

\section{Results for a uniformly distributed growth rate}

In this section, we present detailed results for the probability distribution $p(Z)$ in regimes I and II for a uniformly distributed growth rate, $a \sim \mathcal{U}(b,c)$.

\subsection{Numerical results in regime I (Finite Scale regime)}
\label{sec:num_unif}

\begin{figure*}[htbp]
\includegraphics[width = \textwidth]{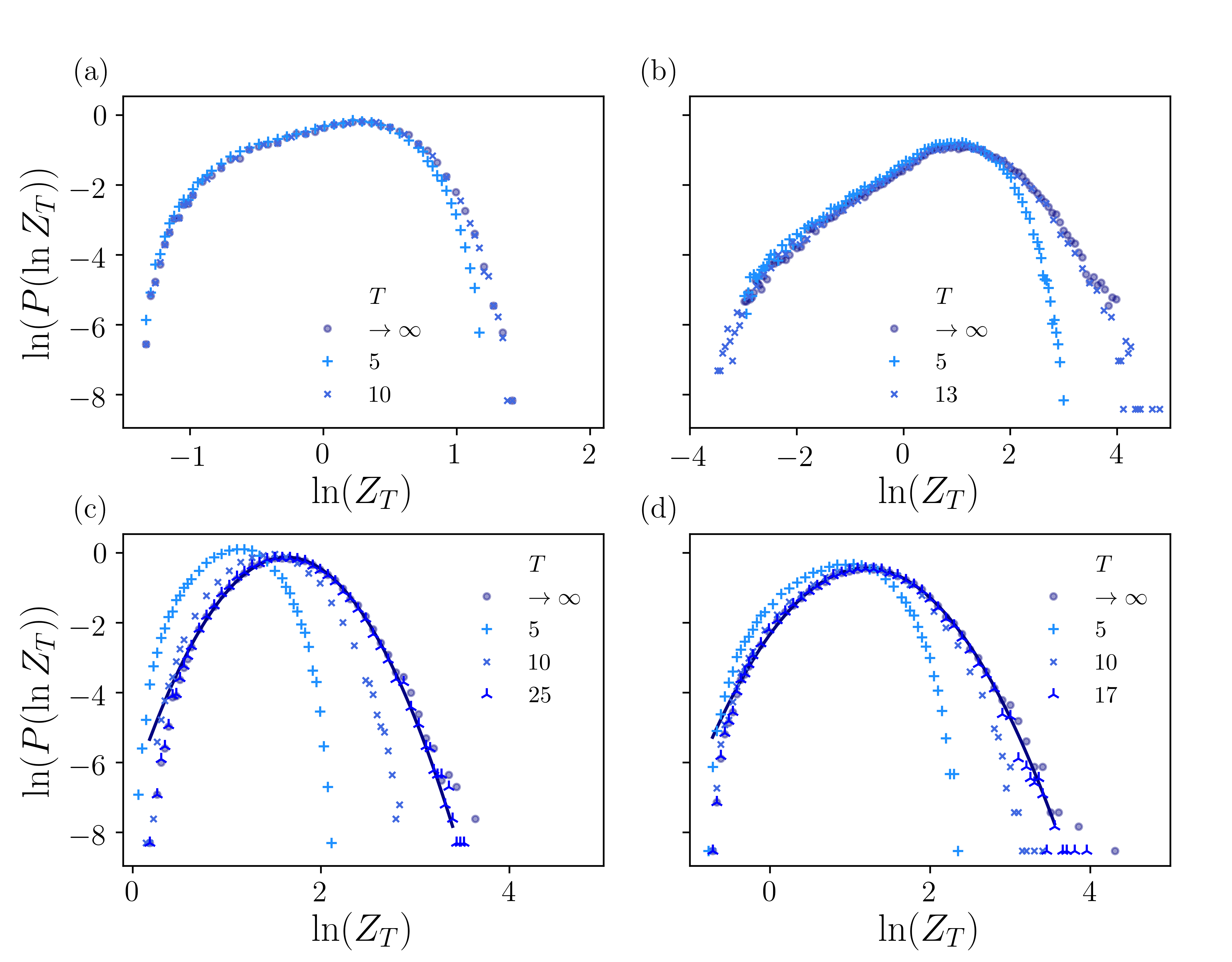}%
\caption{Distribution of the logarithm 
of the sum $Z_T = \sum_{i=1}^{T} z_i  $ for different points in the parameter plane and different $T$ (see legends), until convergence toward the asymptotic distribution of $\ln(Z)$, after time $T_z$. For all the plots the number of realizations for the statistics is $10^5$. The asymptotic distribution, $p(Z)$, is represented by $p(Z_T)$  at time $T = 500$ and the logarithm of the resulting distribution is plotted. Examples are shown with $v_a > \bar{a} -0.7$ (panels (a), $(0.55,0.04)$ and (b),  $(0.8,0.2)$) and with $v_a < \bar{a} -0.7$ (panels (c), $(0.85,0.04)$ and (d), $(0.8,0.08)$). In panel (c) and (d) the maximum  likelihood fit of the asymptotic distribution of $\ln(Z)$ with a log-normal distribution is shown in solid line. 
}\label{fig:finite_size}
\end{figure*}

An exploration of the asymptotic distribution $p(Z)$ for different points $(\bar{a},v_a)$ in the region corresponding to regime I suggests that $p(Z)$ can be approximated by a log-normal distribution for $v_a \lesssim \bar{a}-0.7$ (see Fig.~\ref{fig:finite_size}(c), (d)). In fact, a log-normal approximation is already suitable for the probability distribution $p(Z_T)$ before convergence to the asymptotic distribution $p(Z)$.
 In contrast, $p(Z)$ significantly deviates from a log-normal distribution for $v_a \gtrsim \bar{a}-0.7$ (see Fig.~\ref{fig:finite_size}(a), (b)), and, in these examples, $p(\ln(Z))$ looks like a bimodal distribution, showing a shoulder for values $\ln(Z)<0$.


\subsection{Analytical solution for $p(Z)$ with $a\sim \mathcal{U}(0,c)$ }
\label{sec:analytic_b=0}

In order to fully determine the expression of $p(Z)$ in the particular case where the lower bound of the uniform distribution is $b=0$, it is convenient to write Eq.~\eqref{eq:pZ} in the form of a differential equation. To this end, we first express the asymptotic avalanche size distribution for a uniformly distributed $a$ as follows:
\begin{equation}
\begin{gathered}
 p(Z) = \int_0 ^{+\infty} p(y-1)f_a\tonda{\frac{Z}{y}} \frac{dy}{y}=\\ = \int_0^{+\infty} p(y-1) \frac{\theta\tonda{\frac{Z}{y} - b} - \theta \tonda{\frac{Z}{y} -c }}{c-b}\frac{dy}{y}. \label{eq:integral_unif}
\end{gathered}
\end{equation}
 By using the properties of the Heaviside step function, differentiating with respect to $Z$, and noting that $Z > b$, from Eq.~\eqref{eq:integral_unif} we obtain:
\begin{equation}
p'(Z) = \frac{p\tonda{\frac{Z}{b}-1}}{c-b} \frac{1}{Z} - \frac{p\tonda{\frac{Z}{c}-1}}{c-b} \frac{1}{Z}\theta(Z-c),
\label{eq:diff_eq_supp}
\end{equation}
which is a functional (or delayed) differential equation for $p(Z)$. 

For $b=0$, Eq.~\eqref{eq:diff_eq_supp} reduces to
\begin{equation}
    p'(Z)= -p\tonda{\frac{Z}{c}-1} \frac{1}{c Z} \theta(Z-c).
    \label{diff_eq_unif}
\end{equation}
Here, we used that $\lim_{b \to 0} p(Z/b-1)=0$.
The solution of this equation can be expressed in a piece-wise form as follows:

\begin{equation}
    p(Z)=\begin{cases}
    \hat{p}_1(Z),& Z\in(0,c] \\
    \hat{p}_2(Z),& Z\in(c,2c] \\
    \hat{p}_3(Z),& Z\in(2c,3c] \\
    \dots
    \end{cases}
\end{equation}
In the interval $Z \in (0,c]$, the function $\hat{p}_1(Z)$ is given by
\begin{equation}
\hat{p}_1'(Z) = 0  \Rightarrow \hat{p}_1(Z) = k_1~,
\end{equation}
where $k_1$ is a constant that can be obtained from the normalization of $p(Z)$.
From Eq.~\eqref{diff_eq_unif}, the functions $\{\hat{p}_j(Z),j \geq 2\}$ satisfy the following recurrence relation:
\begin{equation}
    \hat{p}_{j}'(Z)=-\frac{\hat{p}_{j-1}(Z/c-1)}{c Z} \Rightarrow \hat{p}_{j}(Z)=k_{j}-\int\frac{\hat{p}_{j-1}(Z/c-1)}{c Z}\quad j=1,2,\dots
    \label{diff_eq_pj}
\end{equation}
Here, all the constants $\{k_j, j \geq 2\}$ can be expressed in terms of $k_1$ by imposing the continuity of $p(Z)$ at each point $Z=jc$. Then $k_1$ is actually the only normalization constant of the probability $p(Z)$. This is valid for any $c$ such that the product $\prod_{j=2}^T a_j \rightarrow 0$, otherwise $p(Z)$ becomes non-stationary and log-normal, as explained in the main text (for $(\bar{a}, v_a) > \tonda{\bar{a}^{\dag}, v_a^{\dag}}$, along the line $v_a = \bar{a}^2/3$).

For the functions $\{\hat{p}_j(Z), j \geq 2\}$, we present results for the case $c=1$.
In the interval $Z \in (1,2)$, $\hat{p}_2(Z)$ is given by
\begin{equation}
\hat{p}_2'(Z) = -\frac{k_1}{Z}  \Rightarrow p_2(Z) = - k_1 \ln(Z) + k_2~.
\end{equation}
By imposing continuity at $Z=1$ we get $k_2 = k_1$.

The function $\hat{p}_3(Z)$ giving $p(Z)$ in the interval $Z\in (2,3]$ is given by
\begin{equation}
\hat{p}_3(Z) = -k_1\ln(Z)-k_1 \text{Li}_2(Z) +k_1 (\ln(Z-1) - \ln(1-Z))\ln(Z) + k_3~,
\end{equation}
where $\text{Li}_{s+1}$ is the polylogarithm function defined as:
$$\text{Li}_{s+1}(x) = \int_0^x \frac{\text{Li}_s(t)}{t}dt  \qquad \text{and} \qquad	\text{Li}_1(x) = - \ln(1-x)~.$$
The expression for $\hat{p}_4(Z)$ corresponding to the interval $Z\in (3,4]$ can again be expressed in terms of the polylogarithm function:
\begin{equation}
\begin{gathered}
 p(Z) = - k_3 \ln(Z)-k_1 \text{Li}_2(Z) +k_1 (\ln(Z-1) - \ln(1-Z))\ln(Z) + \\
+  k_1 [ -\text{Li}_3(2-Z)  - \text{Li}_3(Z) + \text{Li}_3\tonda{\frac{Z}{2-Z}} -\text{Li}_3\tonda{\frac{Z}{Z-2}}+\\
+\ln\tonda{\frac{Z}{2-Z}} \tonda{ \text{Li}_2\tonda{\frac{Z}{Z-2}} - \text{Li}_2\tonda{\frac{Z}{2-Z}}}+ \text{Li}_2(2-Z)\tonda{\ln(Z) - \ln\tonda{\frac{Z}{2-Z}}} +\\
+ \text{Li}_2(Z-1) \ln(Z) + \text{Li}_2(Z)\tonda{\ln(2-Z) + \ln\tonda{\frac{Z}{2-Z}}} + \\
+\frac{1}{2} \tonda{\ln\tonda{-\frac{2}{Z-2}} + \ln(Z-1) - \ln\tonda{\frac{2(Z-1)}{Z-2}}}\ln^2\tonda{\frac{Z}{2-Z}}+\\
+\tonda{\ln(1-Z) - \ln(Z-1)}\ln(Z)\ln\tonda{\frac{Z}{2-Z}}+\ln(2-Z)\ln(Z-1)\ln(Z)+\\
+\frac{1}{2}\tonda{\ln(Z-1) - \ln(1-Z)}\ln(Z) \tonda{\ln(Z) -2\ln(2-Z)}+\\
\text{Li}_3(1-Z) + \text{Li}_3\tonda{1-\frac{Z}{2}} + \text{Li}_3\tonda{\frac{Z-1}{Z-2}}-\text{Li}_3\tonda{\frac{2(Z-1)}{Z-2}}- \\
+\ln\tonda{\frac{2(Z-1)}{Z-2}}\tonda{\text{Li}_2\tonda{\frac{Z-1}{Z-2}} - \text{Li}_2\tonda{\frac{2(Z-1)}{Z-2}}} +\\
- \text{Li}_2(1-Z)\tonda{\ln(Z-2) + \ln\tonda{\frac{2(Z-1)}{Z-2}}} +\\
- \text{Li}_2\tonda{1-\frac{Z}{2}}\tonda{\ln(Z-1) - \ln\tonda{\frac{2(Z-1)}{Z-2}}}+\\
-\frac{1}{2}\tonda{\ln\tonda{\frac{1}{4-2Z}} +\ln(Z) - \ln\tonda{\frac{Z}{2-Z}} }\ln^2\tonda{\frac{2(Z-1)}{Z-2}} + \\
- \ln(2)\ln(1-Z)\ln\tonda{\frac{2 (Z-1)}{Z-2}}+ \\
+ \frac{1}{2}\ln(2)\ln(1-Z)\tonda{\ln(1-Z) - 2\ln(Z-2)} - \ln(Z-2)\ln(Z-1)\ln \tonda{\frac{Z}{2}} + \\
+ \text{Li}_3(Z) - \text{Li}_2(Z)\ln(-Z) + \frac{1}{2} \tonda{\ln(Z-1)-\ln(1-Z)}\ln^2(-Z)
] + k_4
\end{gathered}
\end{equation}

This shows that the solution becomes quickly complicated, but in principle can be found for all intervals and will involve all degrees of the polylogarithm function.
More pragmatically we can observe that the solution for large $Z$ can be calculated, giving us the asymptotic behavior of $p(Z)$. That is:

\begin{equation}
p'(Z) = - \frac{p(Z)}{Z}  \Rightarrow p(Z) = \frac{k_\infty }{Z} \qquad \text{for} \qquad z \gg 1 .\label{asympt}
\end{equation}

The presented solution is the one plotted in Fig~\ref{fig:numerical_results}(b). 

\subsection{Analytical results in regime II (Power-Law)}

\label{sec:analytic_unif}

We present here the analytical calculations for the case of a uniformly distributed reproduction rate, which was chosen for illustration of the results in the main text. 

For $a \sim \mathcal{U}(b,c)$ a full solution of Eq.~\eqref{eq:pZ} is in general not possible, but still we can obtain a general expression of the power-law exponent $\alpha$ with respect to the model parameters. Indeed, an analytical expression of the function $h(\alpha)$ (see Eq.~\eqref{eq:halpha}) is possible:
 \begin{equation}
 h(\alpha) = \int_0^{+\infty} a^{\alpha-1}\frac{\theta\tonda{a - b} - \theta \tonda{a -c }}{c-b}da -1 = 
  \frac{\quadra{\bar{a} + \sqrt{3 v_a}}^{\alpha} - \quadra{\bar{a} - \sqrt{3 v_a}}^{\alpha}}{2 \alpha \sqrt{3v_a}}-1, \label{eq:def_h(alpha)}
\end{equation}     
where we have injected the relations $b = \bar{a} - \sqrt{3 v_a}$ and $c = \bar{a} + \sqrt{3 v_a}$, and
 written the uniform PDF by means of the Heaviside step function $\theta(x)$, as 
$$f_a(a) = \frac{\theta\tonda{a - b} - \theta \tonda{a -c }}{c-b}$$

The zeros of $h(\alpha)$, if they exist, are the exponent of the power-law tail in region II, thus $\alpha$ is related to $\bar{a}$ and $v_a$ by:
\begin{equation}
-\alpha = \frac{\tonda{\bar{a} - \sqrt{3 v_a}}^\alpha - \tonda{\sqrt{3 v_a} + \bar{a}}^\alpha }{2 \sqrt{3 v_a}}. \label{eq:exp_alpha}
\end{equation}

Eq.~\eqref{eq:exp_alpha} has a trivial solution for $\alpha = 1$, which is not compatible  with the definition of a probability distribution, since it  is not normalizable, besides not being compatible with the numerical observed one (see Fig.~\ref{fig:numerical_results}(c)). 
 Notably,  Eq.~\eqref{eq:exp_alpha} has another solution that can be found numerically and is in very good agreement with exponents of Fig.~\ref{fig:numerical_results}(c). Therefore the asymptotic behavior of $p(Z)$ is proved to follow a power-law tailed distribution, which is an asymptotic solution of Eq.~\eqref{eq:pZ} and whose exponent is determined by the model parameters from Eq.~\eqref{eq:exp_alpha}. 
We are now going to compare these analytical results with the numerical observed ones shown in Fig.~\ref{fig:numerical_results}.  We first note that at the critical line $v_a = 1 -\bar{a}^2$, we find $\alpha = 3 $, which does not depend on  $\bar{a}$ (corresponding to $(\bar{a} \simeq 0.98, v_a \simeq 0.033)$ in Fig.~\ref{fig:numerical_results}(c)). 
 This happens when $m_2 = 1$ and hence, from Eq.~\eqref{varZ_case2}, $\mathrm{Var}(Z_T)$ is divergent with $~T$, consistently with the value $\alpha = 3 $ for which $p(Z)$ has a divergent variance and a convergent mean.  As long as $m_2>1$ and $\bar{a}<1$, we have seen that $\mathrm{Var}(Z_T)$ shows an exponential divergence with $T$, while $\mathbb{E}[Z_T]$ converges. This is the case for the black triangles in Fig~\ref{fig:numerical_results}(c), for which the exponent computed from Eq.~\eqref{eq:exp_alpha} is $\alpha \simeq 2.62$. Another peculiar case is when $\bar{a} = 1$.  Here, Eq.~\eqref{eq:exp_alpha} predicts an exponent $\alpha = 2$ for any possible value of $v_a$. For this value of $\bar{a}$ the average of $Z$  grows linearly with $T$ (see Eq.~\eqref{eq:averageZ_case4}), and Eq.~\eqref{eq:varZ_case4} shows an exponential divergence with $T$. Again, this is consistent with the exponent $\alpha = 2$, for which $p(Z)$ starts having a divergent mean as an effect of the fat tail of its distribution. 
 When both $\bar{a}$ and $m_2$ are larger than $1$ both $\mathbb{E}[Z_T]$ and $\mathrm{Var}(Z_T)$ diverge exponentially with $T$, but as shown in Fig.~\ref{fig:numerical_results}(c) (cyan crosses) the distribution can still be stationary with a very large power-law tail. For that values of parameters, the analytical exponent  $\alpha \simeq 1.29$. 
 
For a uniformly distributed $a$, the boundary between regime II and III given by Eq.~\eqref{eq:Cond_boundary_II_III} of the main text satisfies:
\begin{equation}
\frac{\tonda{\bar{a} + \sqrt{3 v_a}}\tonda{1-\ln(\bar{a} + \sqrt{3 v_a})}}{\tonda{\bar{a} - \sqrt{3 v_a}}\tonda{1-\ln(\bar{a} - \sqrt{3 v_a})}} = 1. \label{eq:critical_curve}
\end{equation}
This defines the curve in the phase diagram above which $p(Z)$ is not stationary anymore and allows us to compute $(\bar{a}^{\dag} \simeq 1.36, v_a^{\dag} \simeq 0.61)$, representing the
largest parameter values for a power-law distribution.

\section{Results for an exponentially distributed growth rate} 
\label{sec:analytic_exp}

\begin{figure}
\centering
\includegraphics[width = 0.7\textwidth]{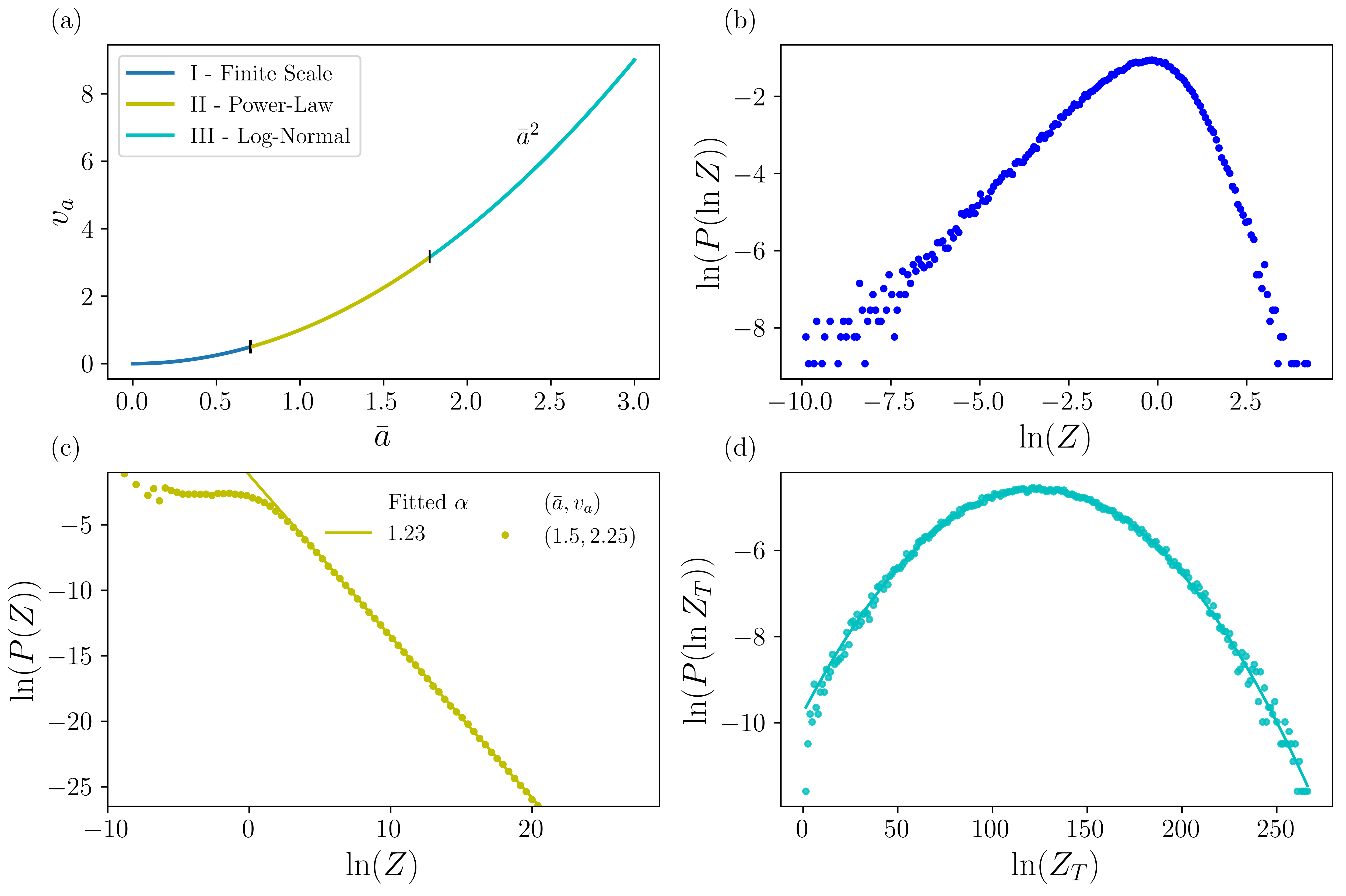}
\caption{ (a) Phase diagram of the distribution of $Z$ for an exponentially distributed reproduction rate, in the model parameter space.  (b) Example of distribution of $p(\ln(Z))$ in regime I, for a reproduction rate $a \sim \text{Exp}(1/0.5)$. (c) Example of distribution of $p(Z)$ in log-log scale for regime II and correspondent power-law fit, for $a \sim \text{Exp}(1/1.5)$. (d) Example of distribution of $p(\ln(Z_T))$ with $T=1000$ in regime III and correspondent fit, for $a \sim \text{Exp}(1/2)$. In plots (b), (c), (d)  symbols (respectively solid lines) are used for numerical results (resp. maximum likelihood fits).  }\label{fig:schema_exp}
\end{figure}

Analytical results are also possible for an exponentially distributed reproduction rate $a \sim \text{Exp}(\bar{a}^{-1})$, with probability density function $f_a(a) = \bar{a}^{-1} e^{- x/\bar{a}}$. In this case there is only one relevant parameter, since $v_a = \bar{a}^2$. This has the consequence that the phase diagram (shown in Fig.~\ref{fig:schema_exp}(a)) is composed of a curve and not of an extended region as for the uniformly distributed growth rate. In Fig.~\ref{fig:schema_exp}(b), (c) and (d) we show a numerical example of avalanche size distribution for each of the three regimes identified in the phase diagram. The three regimes are the same as for a uniformly distributed reproduction rate. 

The expression for $h(\alpha)$ giving the power-law exponent $\alpha$ can be analytically obtained by integration of Eq.~\eqref{eq:halpha}:
\begin{equation}
h(\alpha)  = \bar{a}^{\alpha-1} \Gamma(\alpha) -1,
\end{equation}
where $\Gamma(z) = \int_0^{\infty} x^{z-1}e^{-x} dx$ is the gamma function. The roots of this equation are therefore the exponents of the power-law tail of region II, and can be found numerically. For instance for the example shown in Fig.~\ref{fig:schema_exp}(c) the value of the analytical exponent is $\alpha = 1.23$, which, within statistical error, coincides  with the numerically estimated value given in the figure legend.  Moreover, the curve separating region II and region III can also be expressed from Eq.~\eqref{eq:Cond_boundary_II_III} as:  
\begin{equation}
 \mathbb{E}_a[\ln a] = \gamma + \ln\tonda{\bar{a}} = 0,
\end{equation}
where $\gamma \simeq 0.58$ is the Euler-Mascheroni constant. In this case the point separating the Power-Law and the Log-Normal region is simply 
$ \bar{a} = e^{\gamma}$, while the point $\bar{a} = 1/\sqrt{2}$, corresponding to the condition  $v_a = 1-\bar{a}^2$, separates regions I and II.

\section{Numerical results for a Poisson distributed growth rate}

\begin{figure}
\includegraphics[width = 0.3\textwidth]{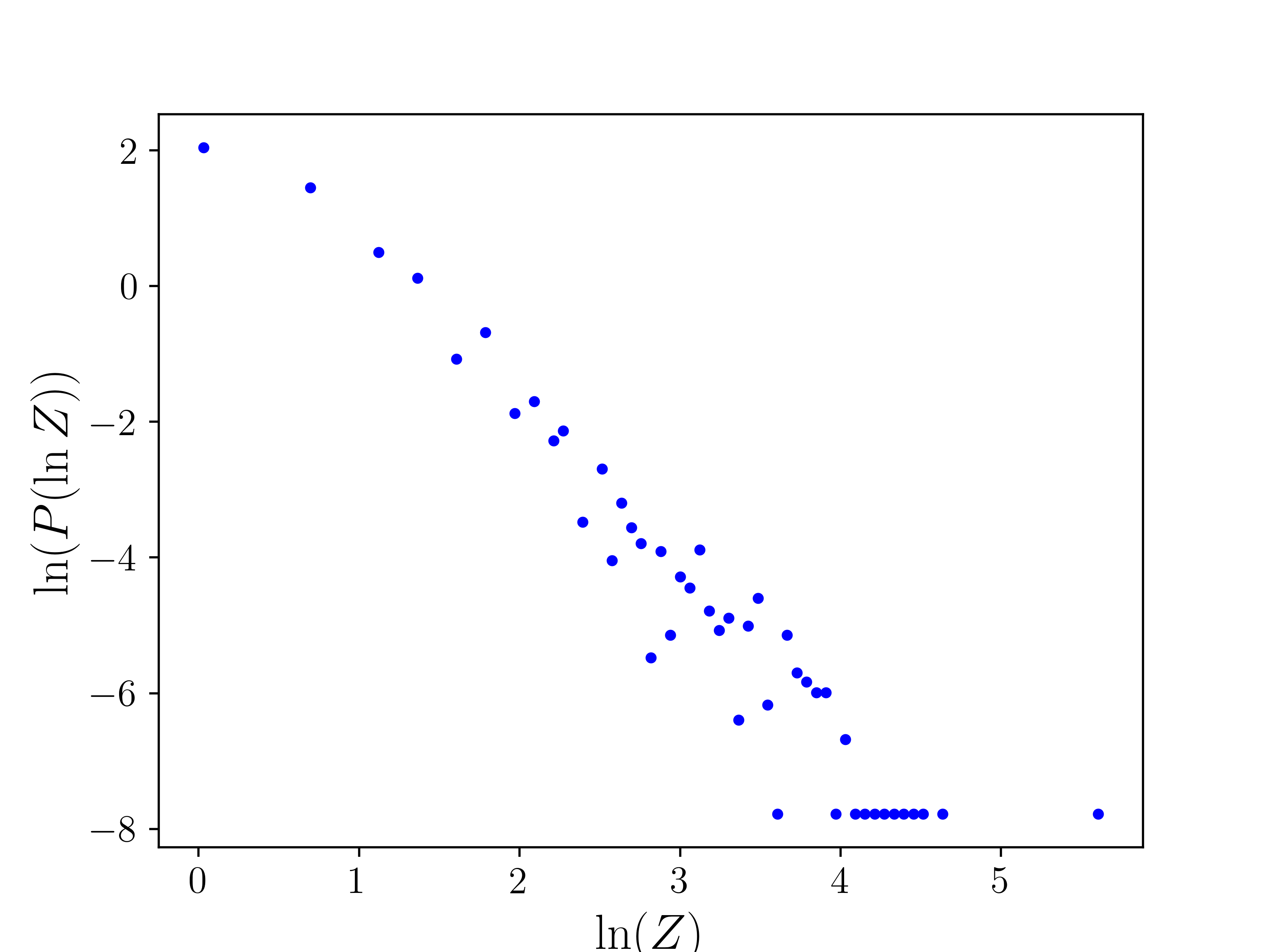}
\includegraphics[width = 0.3\textwidth]{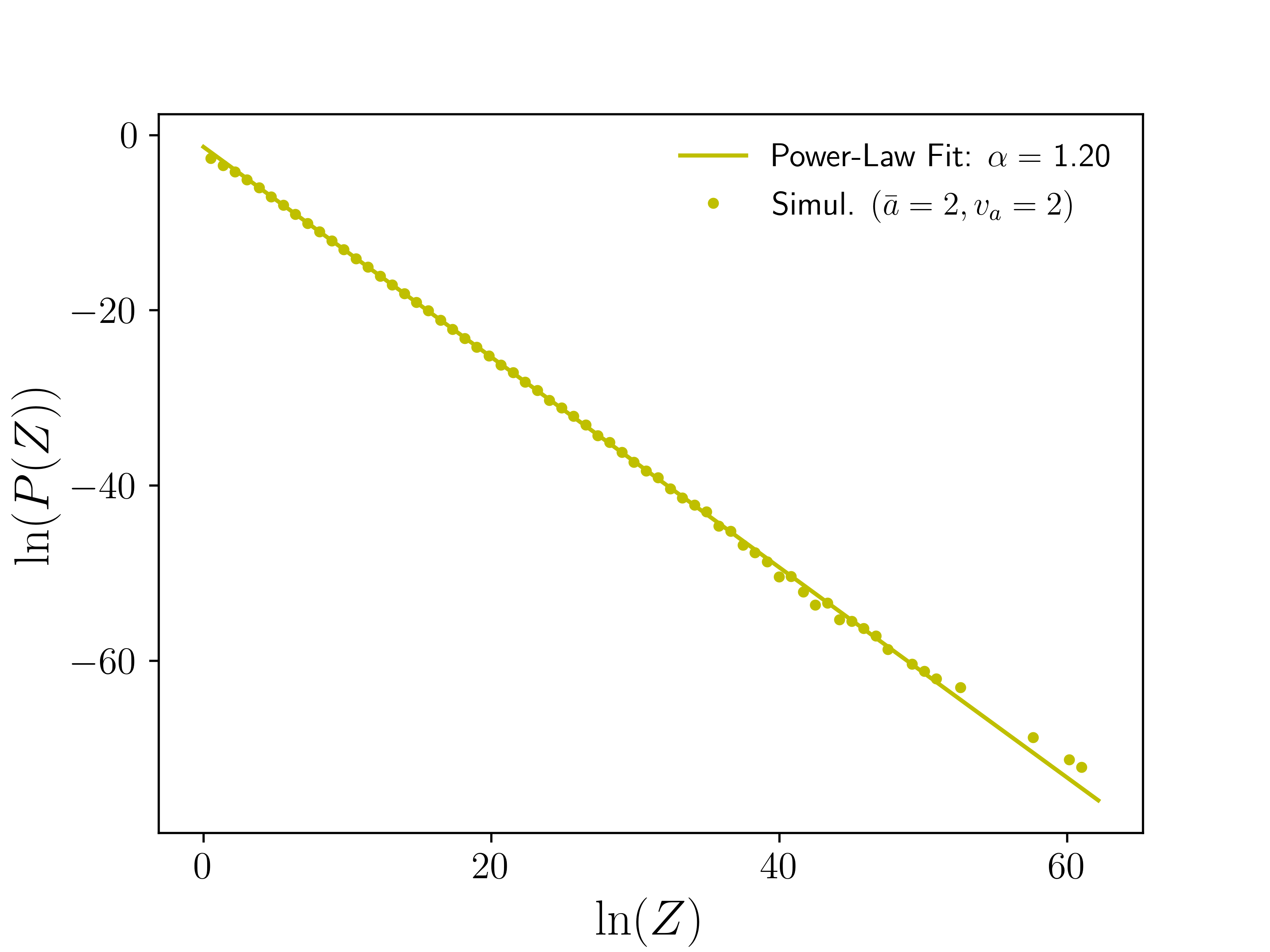}
\includegraphics[width = 0.3\textwidth]{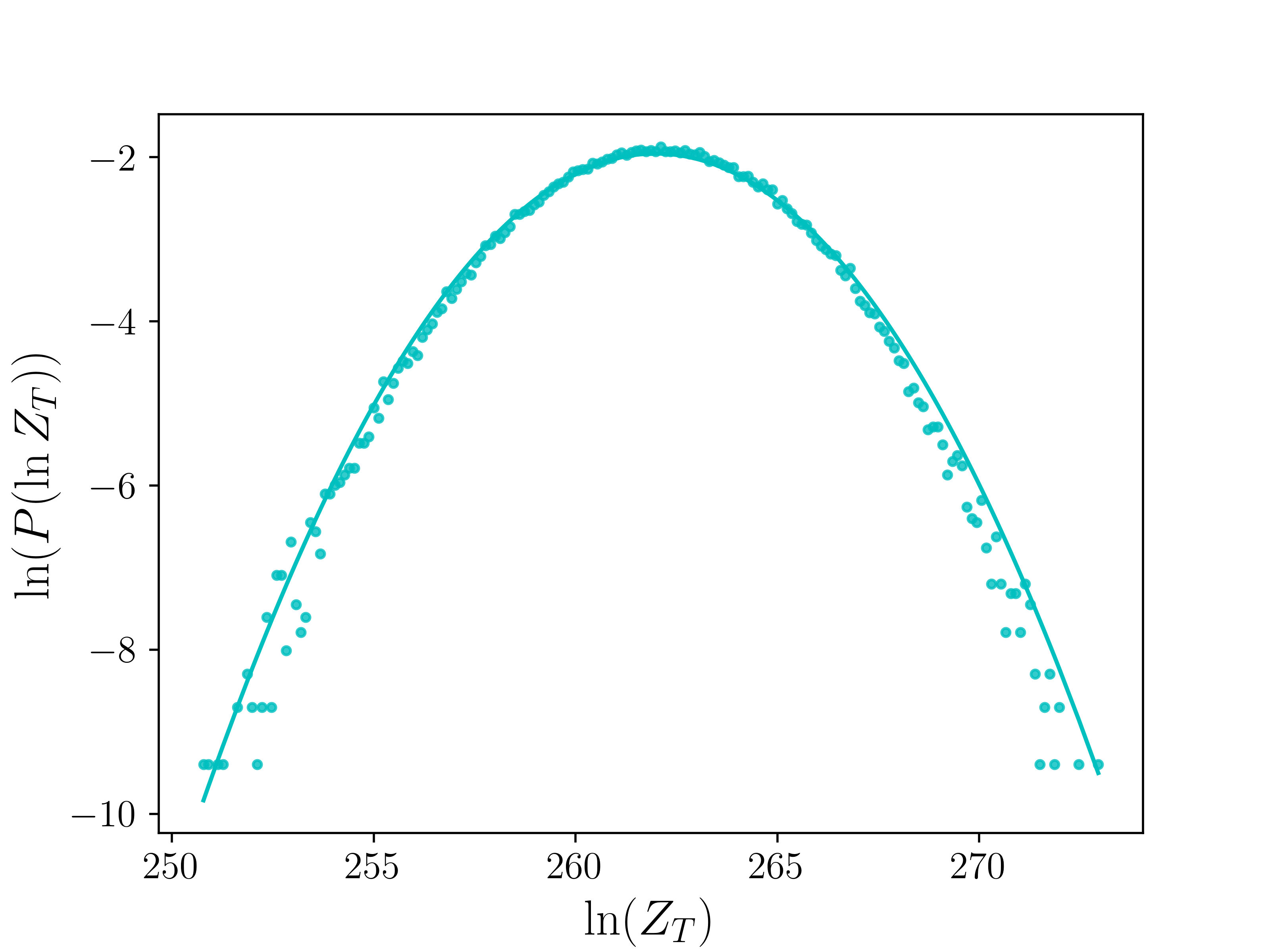}
\caption{\textit{Left:} Example of distribution of the asymptotic $p(\ln(Z))$   in regime I, for a reproduction rate $a \sim \text{Poisson}(0.5)$. \textit{Center:} Example of distribution of the asymptotic $p(Z)$ in log-log scale for regime II and correspondent power-law fit, for $a \sim \text{Poisson}(2)$. \textit{Right:} Example of distribution of $p(\ln(Z_T))$ with $T=100$ in regime III, for $a \sim \text{Poisson}(15)$ and corresponding maximum likelihood fit.}
\label{fig:num_results_poisson}
\end{figure}

In this section we introduce some numerical results for a Poisson distributed growth rate, in order to show that the picture described in the main text is not only true for continuous growth rates, but can also be applied to discrete $a$. In Fig.~\ref{fig:num_results_poisson} we show the same three regimes observed for the uniform and exponential distributions, are also observed for Poisson distributions, where the value $z_i=0$ can be achieved after a finite number of generations. Note that since $Z_T$ here is discrete, in regime I the shape of the asymptotic distribution for $Z$ can be very different. Indeed, with respect for instance to Fig.~\ref{fig:schema_exp}(b) the distribution is cut for $\ln(Z) < 0$. However, the mathematical understanding of the $3$ regimes remains valid. 

 \section{Statistical methods: likelihood estimate of the exponent, likelihood ratio test and bootstrap}
\label{sec:stat_methods} 
 
We describe in this section the statistical techniques used to study the properties of $p(Z_T)$.

\subsection{Maximum likelihood estimate of the exponent and likelihood ratio test}

The mathematical procedure to fit the tail of the distribution is the one described in reference \citep{Clauset2009} and the software used is a lightly  modified version of the python module described in Ref. \citep{Alstott2014} (with some added features, in particular a corrected definition of the cumulative distribution used to perform the fit, defined as $P(X \leq x)$ instead of $P(X < x)$, which can have some impact on the results only for small data-sets, say $<1000$ , that is not our case). Briefly, the fit procedure consists in identifying which is the best interval of the domain to be appropriately fitted by a power-law and in this interval estimate by maximum of likelihood the exponent.

 Moreover, a new fit by maximum of likelihood of the tail  of the distribution with distributions other than the power-law (for instance the log-normal), can be done. We applied this method for instance to estimate the power-law exponent of Fig.~\ref{fig:numerical_results}(c). 
There, the standard deviation of the fit is always smaller than or equal to the last digit, except for the blue circles, where it is $0.03$, meaning that the power-law tail is very well statistically confirmed. Notice that the estimated exponents leads to a diverging mean and/or variance of $Z$.

The likelihood-ratio test \citep{Severini2001} is a test based on the ratio of the likelihood of the model fitted by two different distributions and this can be done to compare a power-law tail with the tails of other distributions.  The null hypothesis of this test is that the likelihood of the $2$ fits (with different distributions) are equal. 
This test was used to check if the tail of the distributions obtained in regime III was log-normally or power-law distributed, showing for all times $T$ of Fig.~\ref{fig:numerical_results}(d) statistical significance (with a significance threshold at $0.05$) that the distribution that was the best fit of the tail was the log-normal. In particular for $T=500$ the p-value of the test was $4 \cdot 10^{-6}$ indicating that the likelihood is significantly better for a log-normal tail.

\subsection{Bootstrap}

Bootstrap was used to assess the suitability of a log-normal distribution for $p(Z_T)$ in regime III. 
   Opposite to other techniques, such as the likelihood ratio test described before,   the goal of the bootstrap technique is to study the statistical information regarding the full distribution, not only the tail. The bootstrap technique is  described for instance in ref. \citep{davison1997}. We developed a python code to apply parametric bootstrap to our case, the bootstrap being parametric since  we are supposing that the distribution is a log-normal and then it depends on $2$ parameters, $\mu_Z$ and $\sigma_Z^2$. Basically, this method aims at checking if the difference between the distribution of $Z_T$ and the log-normal distribution  with parameters $\mu_Z$ and $\sigma_Z$ can be explained  by random fluctuations only.
The algorithm works in the following way:

\begin{itemize}

\item[1 $-$] we do a maximum likelihood estimate of $\mu_Z$ and $\sigma_Z^2$ from the original data (the simulation) and we compute the KS distance, $D^*$ (quantifying the distance between the expected distribution and the one of the data, described just above);

\item[2 $-$]  we generate new random data with the same size as the original one (the simulation), following a log-normal distribution with the estimated parameters $\mu_Z$ and $\sigma_Z^2$;

\item[3 $-$] we estimate by maximum of likelihood the parameters $\mu$ and $\sigma$ of the random data from point $2$ and the distance $D$ between the distribution of the generated random data and the log-normal distribution having as parameters the new $\mu$ and $\sigma$;

\item[4 $-$] we repeat $2$ and $3$  a statistically large number of times (of the order of at least $1000$-$2000$), in order to have a good statistics for the distribution of the distance $D$.
\end{itemize} 

At the end we have a large number of $D$ values and we can thus plot their distribution. Therefore we can compute the probability  that the $D$ obtained with the bootstrap process   is bigger than the $D^*$ of the simulation. This gives us a p-value under the null hypothesis: $Z_T$ follows a log-normal distribution. This p-value  then tells us the probability that $Z$ follows a log-normal. With a typical significance threshold of $5\%$, we can then draw a conclusion about the null hypothesis.

For this study, we developed a code to use this technique, obtaining strong statistical evidence that the difference between the $3$ distributions of Fig~\ref{fig:numerical_results}(d) and the maximum likelihood fit of the log-normal can be explained by random fluctuations only. For the parameters in Fig~\ref{fig:numerical_results}(d) the p-value$>0.05$ for all $T$ and increases with $T$ up to $0.31$, for $T = 500$. We can thus draw the conclusion that the null hypothesis of zero distance between the log-normal and the distribution of $Z_T$ can be accepted for all the $T$ considered.
Moreover, this suggests a convergence of $p(Z_T)$ toward a log-normal with $T$, as expected from the Central Limit Theorem. 

 Finally, we found that the p-value 
remains quite stable when increasing the number of repetitions, $n_{rep}$.  This means that even the tails of the distribution are well fitted by the log-normal distribution, since, by increasing $n_{rep}$, events with probability of the order of $1/n_{rep}$ (and therefore very rare) become accessible to the simulation. We did this by checking the p-value for $T=500$ and $n_{rep} = 10^3,10^5,10^5, 10^6$, where the last value is the one chosen for Fig.\ref{fig:numerical_results}(d).

\section{Growth rate distribution for branching processes}
\label{sec:branching}

\begin{figure}
\centering
\includegraphics[width = 0.4\textwidth]{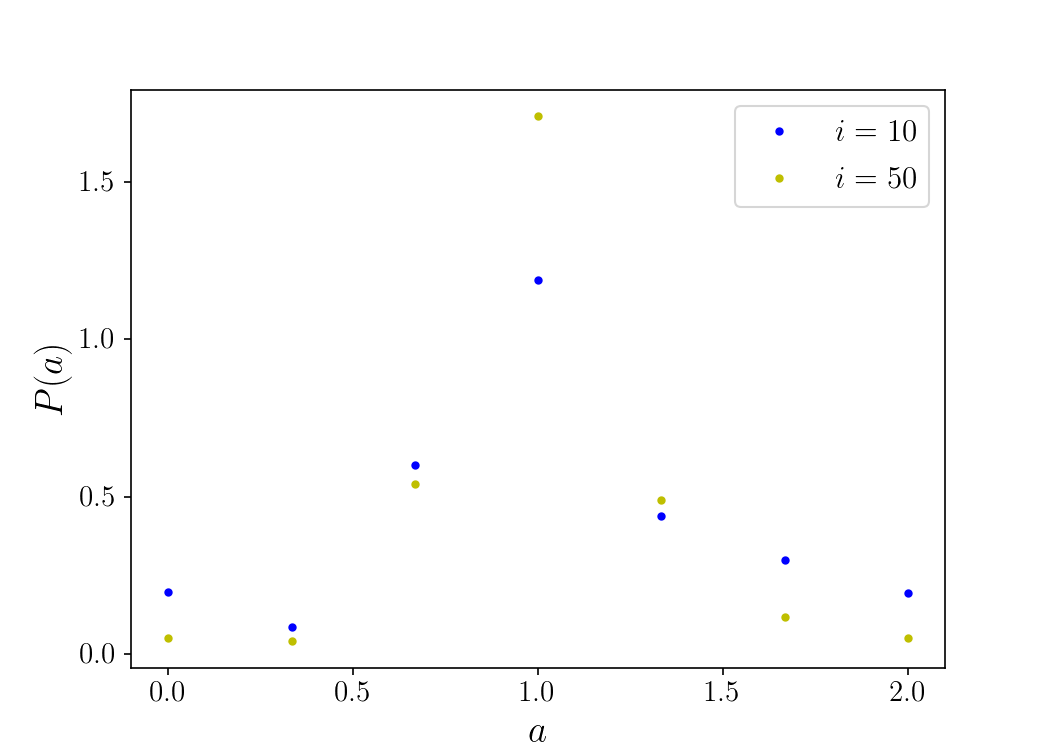}
\caption{Reproduction rate distribution of a critical branching process with $p=0.5$ and $n=2$,  for $i = 10, 50$, from numerical simulations.}\label{fig:branching}
\end{figure}

\begin{figure}
\centering
\includegraphics[width = 0.4\textwidth]{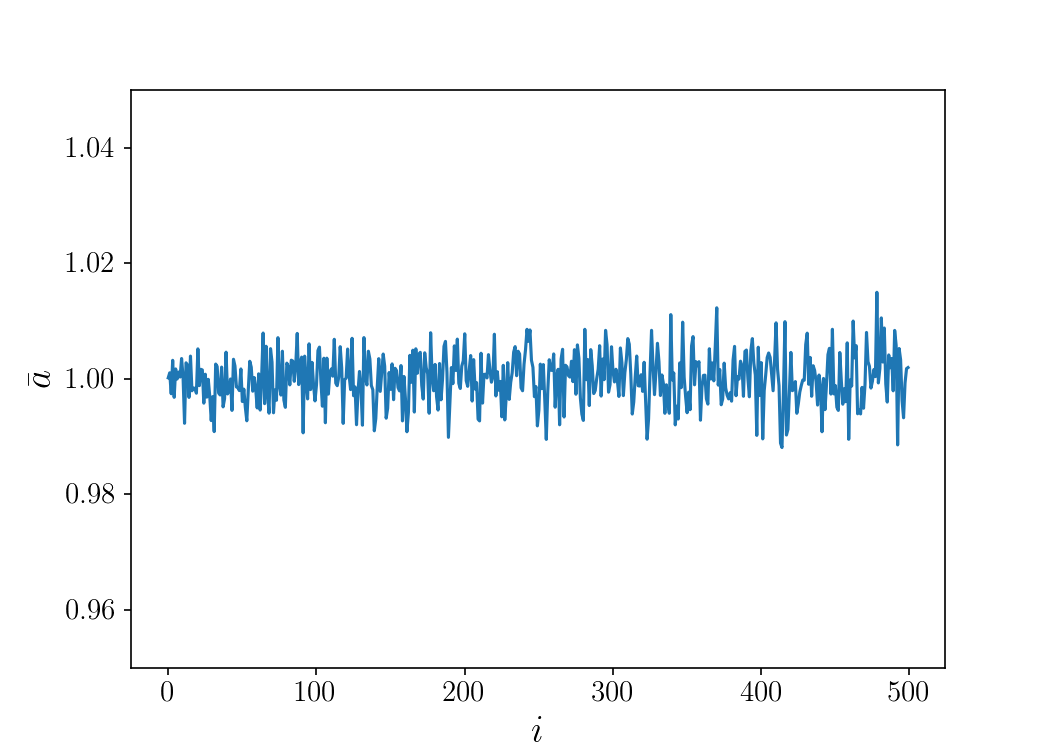}
\includegraphics[width = 0.4\textwidth]{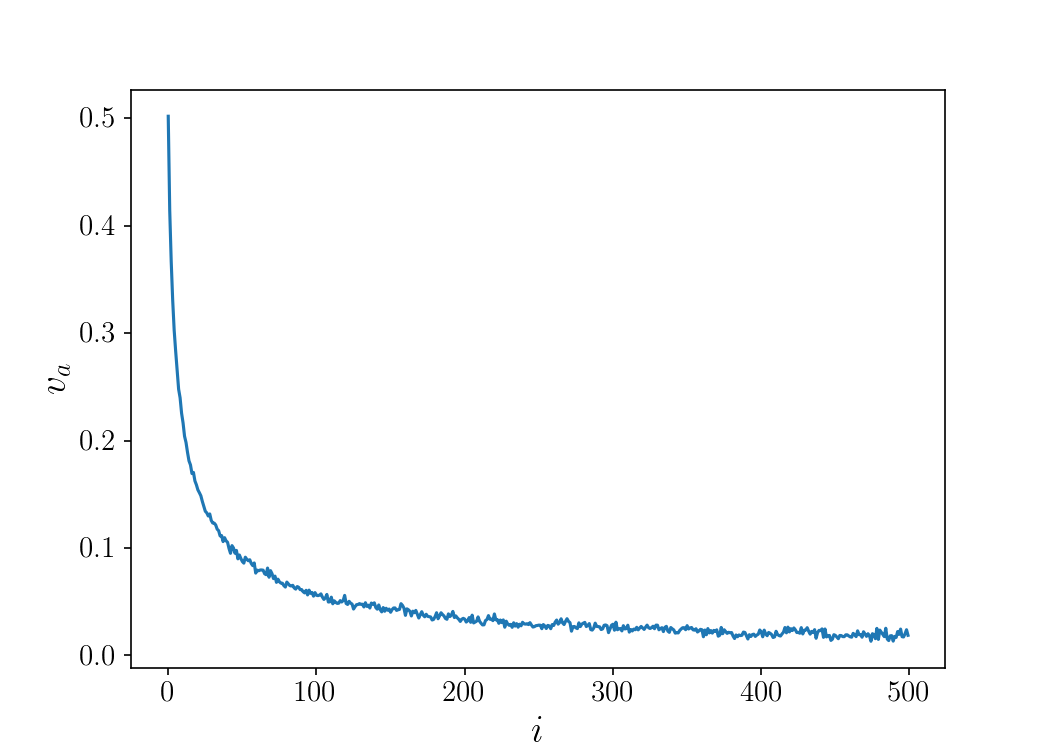}
\caption{Behavior of the mean $\bar{a}$ (left) and the variance $v_a$ (right) with time $i$ for the critical branching process with $p=0.5$ and $n=2$, from numerical simulations. }\label{fig:branching_a_bar_var_a}
\end{figure}

The size $z_{i+1}$ of the population in a branching process at generation $i+1$ is given by the sum of offspring $\{k_j,\, j=1,2,\dots,z_i\}$ of the $z_i$ individuals in generation $t$, i.e. $z_{i+1}=\sum_{j=1}^{z_i} k_j$ \citep{Harris1963,Corral2013}. The branching process can be mapped to the multiplicative process given by Eq.~\eqref{eq:process_def} if we define a growth rate 
\begin{equation}
    a_i=\frac{\sum_{j=1}^{z_i} k_j}{z_i}~.\label{eq:a_branching}
\end{equation}
In particular, we studied the probability distribution of  $\{a_i\}$ for branching processes in which the number of offspring obeys a binomial distribution, $k \sim B(n,p)$.  Fig.~\ref{fig:branching} shows examples of the distribution of $a_i$ for a critical branching process with $n=2$ and $p=1/2$. As expected, $\bar{a} \simeq np =1$. The variance $v_a$ decreases to $0$ with $i$, as shown in Fig.~\ref{fig:branching_a_bar_var_a}. The distribution shown in Fig.~\ref{fig:branching} is discrete, because the in Eq.~\eqref{eq:a_branching} $z_i$ remains always of the order of $1$, and therefore $a_i$ can only take sparse rational values.

  Besides this, in branching processes the two parameters $\bar{a}$ and $v_a$ cannot be independently tuned. For instance, for a critical branching process $\bar{a} = 1$ and only the initial value of $v_a = 1-p$ can be partially tuned, but it still has to verify $p = 1/n$. Also, the dynamics of the $2$ parameters is fixed by the rules governing branching processes.  
  
  This is a particular case of our model, which leads to the known avalanche size distribution of branching processes, that, since they are discrete processes,  can be computed via the generating function method \citep{Harris1963,Corral2013}.



\bibliographystyle{apsrev4-2} 
\bibliography{biblio_mendeley}